\documentclass[a4paper,12pt,tightenlines,notitlepage,nofootinbib]{revtex4-2}
\usepackage{amsmath,amssymb,setspace}
\usepackage{graphicx}
\usepackage{newtxtext}
\usepackage{newtxmath}
\usepackage[normalem]{ulem}
\usepackage[nodayofweek,ddmmyy]{datetime}
\usepackage[x11names]{xcolor}
\usepackage[unicode,colorlinks,citecolor=Blue3,linkcolor=Blue3,bookmarks=true]{hyperref}

\newcommand{\eg}{{\it e.g.\,}}
\newcommand{\const}{\mathop{\rm const}\nolimits}
\newcommand{\svector}[2]{\begin{pmatrix}#1 \\ #2 \end{pmatrix}}
\newcommand{\sh}{\mathop{\rm sinh}\nolimits}
\newcommand{\ch}{\mathop{\rm cosh}\nolimits}

\DeclareSymbolFont{bbold}{U}{bbold}{m}{n}
\DeclareSymbolFontAlphabet{\mathbbold}{bbold}

\begin{document}



\title{A new double-pass type of the optical spring}


\author{F.Ya.Khalili}
\email{farit.khalili@gmail.com}
\affiliation{Russian Quantum Center, Skolkovo IC, Bolshoy Bulvar 30, bld.\ 1, Moscow, 121205, Russia}

\begin{abstract}
  In detuned optical cavities, the radiation pressure force acting on the mirrors depends on their displacements. This is equivalent to the rigidity (the optical spring), inserted between the mirrors. This effect can be used for optimization of the mechanical susceptibility of probe mirrors in high-precision force sensors. However, in some cases, the use of detuned cavities or even just any high-finesse cavities could be problematic due to technological constraints.

  We consider a new type of the optical spring that does not require the cavity (but can use a resonance tuned one to increase the optomechanical coupling). Instead, it uses the double interaction of the probing light with the mechanical object. We propose two possible implementation of this concept, suitable, respectively, for the atomic spin ensembles and for the laser gravitational-wave detectors.
\end{abstract}

\maketitle


\section{Introduction}

In opto- and electro-mechanical systems, the radiation pressure force $F_{\rm R.P.}$ acting on the mechanical object could depend on the position $x$ of this object. The linear in $x$ component of this dependence is known as the optical, or, in the more general context, electromagnetic (e.m.) rigidity. The canonical example is a cavity with the movable mirror, pumped at the frequency $\omega_p$ detuned from the cavity eigen frequency $\omega_o$ (see Fig.\,\ref{fig:std}). In this case, mechanism of the e.m. rigidity can be explained as follows: (i) the movable mirror introduces the phase shift into the reflected light, proportional to the mirror position $x$; (ii) the cavity detuning couples the phase and amplitude quadratures of the light in the cavity, making the amplitude quadrature $x$-dependent as well; (iii) this means that the radiation pressure force acting on the cavity mirrors now depends on $x$.

Note also that the radiation pressure follows the changes of $x$ with a lag proportional to the relaxation time of the cavity. Therefore, the rigidity (the real part of the spring constant) is accompanied by damping (the imaginary part). Both these effects were first observed in 1964 in a radio-frequency system \cite{64a1eBrMi} and later in microwave \cite{70a1eBrMaTi} and optical \cite{Dorsel_PRL_51_1550_1983} resonators.

According to the Fluctuation-Dissipation theorem, any damping is accompanied by the noise.  However, the noise temperature of the e.m. damping could be made very low, approaching the ground state level \cite{01a1BrKhVo, 01a2Kh, 02a1BrVy}. Therefore, the e.m. rigidity and damping can be used for shaping response functions of mechanical objects without introducing significant additional noise. This possibility was exploited successfully in numerous experiments aimed at the preparation of mechanical resonators in the quantum state close to the ground one, see the review \cite{Aspelmeyer_RMP_86_1391_2014}.

Another promising potential application of the optical rigidity is the optimization of response functions of the optomechanical sensors of small forces, in particular, of the laser gravitational-wave (GW) detectors. Sensitivity of these devices is limited by the interplay of two kinds of the quantum noises: the measurement imprecision noise and the back action force. In the optical sensors case, they originate from the phase and the power fluctuations of the probing light, respectively. Their balance corresponds to the so-called Standard Quantum Limit (SQL) \cite{67a1eBr, 92BookBrKh}, inversely proportional to the mechanical susceptibility of the probe object, see Eq.\,\eqref{S_SQL}. In the laser GW detectors, the probe objects can be considered as almost free masses. By adding the low-noise optical spring to the free mass, it can be converted to a harmonic oscillator, highly sensitive to the near-resonance signals, thus reducing the SQL in the frequency band around the mechanical resonance frequency \cite{01a1BrKhVo, 01a2Kh, 12a1DaKh}. Moreover, it is known that in the very long cavities of the GW detectors, the optical rigidity has a sophisticated frequency dependence, which allow to significantly broaden the high-sensitivity band \cite{Buonanno2002, 11a1KhDaMuMiChZh, 12a1DaKh, 19a1DaKhMi}.

The key requirements for implementation of the ``canonical'' type of the e.m. rigidity discussed above are (i) the use of a cavity, which (ii) has to be detuned from the carrier frequency. However, there are important cases where fulfillment of these requirements could be challenging. In particular, the optical rigidity is not used in the contemporary (second generation) GW detectors. The main reason for this is the technical hassles associated with the implementation of detuned operation in large-scale interferometers with suspended mirrors (the use of the optical rigidity is considered for the third generation GW detector Einstein Telescope that presumably will start operating around 2035 \cite{Hild_CQG_28_094013_2011}).

Another example is adjusting of response functions of atomic spin ensembles (the spin oscillators) \cite{Duan_PRL_85_5643_2000, Julsgaard_Nature_413_400_2001, Hammerer_RMP_82_1041_2010}. In presence of an external constant magnetic field, the collective spin of such a system behaves similar to an ordinary harmonic oscillator which the eigen frequency equal to the Larmor precession frequency. At the same time, the spin oscillator eigen frequency could be made both positive or negative, depending on the direction of the magnetic field relative to the sum angular momentum of the spin ensemble. It is known \cite{Polzik_AnnPhys_527_A15_2014}, that combining an optomechanical sensor with the auxiliary negative-frequency spin system, it is possible to cancel the back action noise and obtain the sensitivity overcoming the Standard Quantum Limit, see details in Sec.\,\ref{sec:hybrid}. The necessary condition for this is the anti-symmetry of the main and auxiliary systems: their optomechanical coupling rates and the absolute values of the susceptibilities have to be equal to each other, while the signs of the susceptibilities have to be opposite. This concept was demonstrated experimentally in Ref.\,\cite{Moeller_Nature_547_191_2017} using the drum mode of a dielectric membrane with the eigen frequency close to 1\,Mhz.

The mechanical eigen frequencies in the laser GW detectors are much smaller, about 1\,Hz. Unfortunately, the corresponding decrease of the spin oscillator eigen frequency entails the undesirable increase of its damping rate due to the so-called {\it power broadening} effect \cite{18a1KhPo, 19a1ZePoKh}. In Ref.\,\cite{19a1ZePoKh} it was proposed to use the {\it virtual rigidity} method \cite{95a1VyZu, 12a1DaKh} to reduce the apparent eigen frequency of the spin oscillator by introducing cross-correlation between the imprecision noise and the back action one. It is known, however, that quantum correlations are vulnerable to losses. Therefore, the use of the physical optical rigidity could be preferable (see Sec.\,\ref{sec:losses}); alternatively, both methods could be used together. However, the high-finesse cavity, required for the canonical optical rigidity, can not be used with spin ensembles due to the optical losses introduced by the  spin-protecting coating of the atomic cells.

In this paper, we consider another type of the optical rigidity, which does not require the cavity (but can use a resonance tuned one to increase the optomechanical coupling). Instead, it uses the sequence of two interactions of the probing light with the mechanical object in order to introduce the $x$-dependence into amplitude quadrature of the light. Various implementations of this concept are possible, including the one based on the same optical topology as the two-pass quantum speedmeter interferometer proposed in Ref.\,\cite{18a1DaKnVoKhGrStHeHi}.

The paper is organized as follows. In the next section, we remind the reader the main equations describing the quantum optomechanical force sensors. Then in Sec.\,\ref{sec:double} we introduce the idea of the double interaction e.m. rigidity and discuss its possible implementations. In Sec.\,\ref{sec:losses} we show that the real ``hardware'' rigidity is more robust to the detection losses than the virtual one. Finally, in Sec.\,\ref{sec:conclusion}, we summarize the results of this work.

\section{Basic optomechanical system}\label{sec:basic}

\begin{figure}
  \includegraphics[scale=1.2]{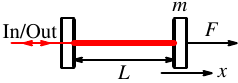}
  \caption{The basic optomechanical system}\label{fig:std}
\end{figure}

Let us start with the basic optomechanical setup. consisting of an optical mode which eigen frequency depends on the position $x$ of the mechanical object $m$:
\begin{equation}
  \omega(x) = \omega_o\biggl(1-\frac{x}{L}\biggr) ,
\end{equation}
where $\omega_o$ is the unperturbed eigen frequency of the mode and $1/L$ is the coupling factor. A simple example using a Fabry-Perot cavity is shown in Fig.\,\ref{fig:std}. Here  the left mirror is partly transparent, providing the coupling of the intracavity field with the external world, the right mirror (the mechanical object $m$) is perfectly reflective and movable, $x$ is its displacement from the initial position, $F$ is the external classical force that has to be detected, and $L$ is the unperturbed distance between the mirrors. It is known that other more sophisticated optomechanical schemes, in particular the Michelson/Fabry-Perot topology of the laser GW detectors, can be reduced to this simple one with some effective parameters \cite{Buonanno2003, 12a1DaKh, 16a1DaKh}.

Throughout this paper, we use the standard approach, based on the two-photon formalism \cite{Caves1985, Schumaker1985}, and the linearized Heisenberg–Langevin equations with the strong classical (mean value) components of the optical fields detached from the quantum fluctuations \cite{12a1DaKh}. We suppose that the cavity is excited by a strong classical light that is tuned in resonance with the frequency $\omega_o$. For simplicity, we assume the bad-cavity approximation here:
\begin{equation}\label{bad^cavity}
  |\Omega| \ll \gamma \,,
\end{equation}
where $\Omega$ is the running (sideband) frequency and $\gamma$ is the cavity half-bandwidth. Note that in the contemporary GW detectors, the mechanical resonance frequencies, achievable by means of the optical rigidity, well satisfy the inequality \eqref{bad^cavity} due to the optical power limitations. This inequality is satisfied also in the spin oscillator based setups. It have to be noted however that using the bad cavity approximation, we neglect the optical damping here. We are planning to consider it the continuation paper.

In this case, equations of motion of our system can be presented as follows:
\begin{equation}\label{IO_10}
  \svector{B^c}{B^s} = \svector{A^c}{A^s}
  = \sqrt{\frac{2I_0}{\hbar\omega_o}}\svector{\cos\phi}{\sin\phi} ,
\end{equation}
\begin{subequations}\label{IO_11}
  \begin{gather}
    \svector{\hat{b}^c(\Omega)}{\hat{b}^s(\Omega)}
      = \svector{\hat{a}^c(\Omega)}{\hat{a}^s(\Omega)}
        + \Upsilon\svector{-\sin\phi}{\cos\phi}\hat{X}(\Omega) \,, \\
    \chi^{-1}(\Omega)\hat{X}(\Omega) = f(\Omega)
      + \Upsilon(\cos\phi\ \sin\phi)\svector{\hat{a}^c(\Omega)}{\hat{a}^s(\Omega)} .
      \label{IO(b)}
  \end{gather}
\end{subequations}
Here $A^{c,s}$ and $B^{c,s}$ are the classical (mean) values of the cosine and sine quadratures of, respectively, the input and output optical fields, $\hat{a}^{c,s}(\Omega)$ and $\hat{b}^{c,s}(\Omega)$ are the corresponding quantum two-photon quadratures, satisfying the commutation relations
\begin{equation}
  [\hat{a}^c(\Omega), \hat{a}^s(\Omega')] = [\hat{b}^c(\Omega), \hat{b}^s(\Omega')]
    = 2\pi i\delta(\Omega-\Omega') \,,
\end{equation}
$I_0$ is the mean incident optical power,  $\phi$ is the phase of the classical field,
\begin{gather}
  X = \frac{x}{\sqrt{\hbar/m}}
  \intertext{and}
  f = \frac{F}{\sqrt{\hbar m}}
\end{gather}
are the normalized mechanical position and force, and $\chi$ is the massless mechanical susceptibility.

The factor
\begin{equation}
  \Upsilon = \frac{1}{\gamma L}\sqrt{\frac{8\omega_oI_0}{m}}
\end{equation}
describes the optomechanical coupling rate. Up to the redefinition of this factor, Eqs.\,\eqref{IO_10},\eqref{IO_11} are applicable also to the cavity-less single movable mirror case, as well as to the atomic spin system setups. In the former case,
\begin{equation}
  \Upsilon = \frac{1}{c}\sqrt{\frac{8\omega_oI_0}{m}} \,,
\end{equation}
In the latter one,
\begin{equation}
  \Upsilon \propto \sqrt{\Gamma_S\Omega_S} \,,
\end{equation}
where $\Omega_S$ is the spins Larmor frequency, and $\Gamma_S$ is the coupling factor proportional to the optical power probing the spin system
\cite{Moeller_Nature_547_191_2017, 19a1ZePoKh}.

In order to provide the baseline for our consideration, start with the case of the {\it simple position meter} \cite{12a1DaKh}. Suppose first without loss of generality that the sine quadrature $\hat{b}^s$ is measured by the homodyne detector. Then suppose also that $\phi=0$. In this case, combining Eqs.\,\eqref{IO_11}, we obtain that
\begin{equation}
  \hat{b}^s(\Omega) = \Upsilon\chi(\Omega)[f(\Omega) + \hat{f}_{\rm sum}(\Omega)] ,
\end{equation}
where
\begin{equation}\label{f_sum_SQL}
  \hat{f}_{\rm sum} (\Omega)
  = \frac{\chi^{-1}(\Omega)}{\Upsilon}\hat{a}^s(\Omega) + \Upsilon\hat{a}^c(\Omega)
\end{equation}
is the normalized sum quantum noise. Finally, suppose that the input fields is prepared in the ground state. In this case, the double-sided spectral densities of the quadratures $\hat{a}_{c,s}$ are equal to $1/2$ and the spectral density of $\hat{f}_{\rm sum}$ is equal to
\begin{equation}\label{S_sum_SQL}
  S(\Omega) = \frac{1}{2}\biggl[\frac{|\chi(\Omega)|^{-2}}{\Upsilon^2} + \Upsilon^2\biggr]
  \ge S_{\rm SQL}(\Omega) \,,
\end{equation}
where
\begin{equation}\label{S_SQL}
  S_{\rm SQL}(\Omega) = |\chi(\Omega)|^{-1}
\end{equation}
is the force-normalized form of the Standard Quantum Limit.

\section{Double-interaction optical spring}\label{sec:double}

\begin{figure}
  \includegraphics[scale=1.2]{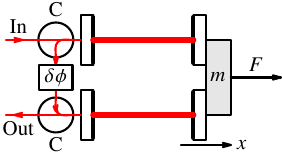}
  \caption{The concept of the double-pass optomechanical system. C: optical circulators. The block labeled ``$\delta\phi$'' introduces a given phase shift between the classical and and quantum components of the probing field. The end mirrors of the both cavities are rigidly attached to the mechanical object $m$. }\label{fig:double}
\end{figure}

Let us consider now the double pass setup sketched in Fig.\,\ref{fig:double}. Here the probing light interact successively with two identical optical modes coupled to the same probe mass. After the first interaction, the {\it phase quadrature} of the probing light carries out information on the mechanical position $x$. Between the interactions, some given phase shift $\delta\phi$ between the classical and and quantum components of the probing field is introduced. It plays the same role as the detuning in the canonical optical rigidity case, making the {\it amplitude quadrature} of the light also dependent on $x$. The practical means how this phase shift can be implemented are discussed in the end of this section. Here, just to be specific, we assume that the classical carrier phase $\phi$ is modified in some way by the block labeled ``$\delta\phi$''.

We enumerate the optical modes by the subscript $i=1,2$ and, up to this subscript, reuse Eqs.\,\,\eqref{IO_10},\,\eqref{IO_11} and the notations therein. As a result, we obtain the following equations:
\begin{subequations}\label{IO_20}
  \begin{gather}
    \svector{B_i^c}{B_i^s} = \svector{A_i^c}{A_i^s}
      = \sqrt{\frac{2I_0}{\hbar\omega_o}}\svector{\cos\phi_i}{\sin\phi_i} , \\
    \svector{A_2^c}{A_2^s} = \mathbb{R}\svector{A_1^c}{A_1^s} ,
  \end{gather}
\end{subequations}
\begin{subequations}\label{IO_21}
  \begin{gather}
    \svector{\hat{b}_i^c(\Omega)}{\hat{b}_i^s(\Omega)}
      = \svector{\hat{a}_i^c(\Omega)}{\hat{a}_i^s(\Omega)}
        + \Upsilon\svector{-\sin\phi_i}{\cos\phi_i}\hat{X}(\Omega) \,, \\
    \svector{\hat{a}_2^c(\Omega)}{\hat{a}_2^s(\Omega)}
      = \svector{\hat{b}_1^c(\Omega)}{\hat{b}_1^s(\Omega)} , \\
    \chi^{-1}(\Omega)\hat{X}(\Omega) = f(\Omega)
      + \Upsilon\sum_{i}(\cos\phi_i\ \sin\phi_i)
          \svector{\hat{a}_i^c(\Omega)}{\hat{a}_i^s(\Omega)} .
  \end{gather}
\end{subequations}
where $\mathbb{R}$ is the unitary matrix rotating the classical amplitudes vector by the angle $\delta\phi = \phi_2-\phi_1$.

Introduce the following notation:
\begin{equation}
  \phi_{1,2} = \phi \pm \psi \,.
\end{equation}
Combining Eqs.\,\eqref{IO_21} and renaming for brevity and for uniformity with our subsequent consideration
\begin{equation}
  \hat{a}_1^{c,s} \to \hat{a}^{c,s} \,, \quad \hat{b}_2^{c,s} \to \hat{b}^{c,s} \,,
\end{equation}
we obtain:
\begin{subequations}\label{IO_K}
  \begin{gather}
    \svector{\hat{b}^c(\Omega)}{\hat{b}^s(\Omega)}
      = \svector{\hat{a}^c(\Omega)}{\hat{a}^s(\Omega)}
        + \Upsilon_\kappa\svector{-\sin\phi}{\cos\phi}\hat{X}(\Omega) \,, \\
    \chi^{-1}(\Omega)\hat{X}(\Omega) = f(\Omega)
      + \Upsilon_\kappa(\cos\phi\ \sin\phi)
          \svector{\hat{a}^c(\Omega)}{\hat{a}^s(\Omega)}
      - \kappa\hat{X}(\Omega) \,, \label{IO_K(b)}
  \end{gather}
\end{subequations}
where
\begin{equation}\label{Ups_eff}
  \Upsilon_\kappa = 2\Upsilon\cos\psi
\end{equation}
is the effective coupling factor, and
\begin{equation}\label{kappa}
  \kappa = \frac{K}{m} = \Upsilon^2\sin2\psi = \frac{\Upsilon_\kappa^2}{2}\tan\psi \,.
\end{equation}
The last term in Eq.\,\eqref{IO_K(b)} describes the component of the radiation pressure force, proportional to the mechanical coordinate $X$ --- that is, the optical rigidity, with $\kappa$ being the massless spring factor.

\begin{figure}
  \includegraphics{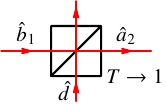}
  \caption{Modification of the classical component of the optical beam using the asymmetric beamsplitter.}\label{fig:comb_simple}
\end{figure}

The optomechanical scheme equivalent to the one shown in Fig.\,\ref{fig:double} was analyzed also in the work \cite{Markowitz_2023}. While the goal of that research was different, the numerical simulation done by the author of \cite{Markowitz_2023} showed that the optical spring indeed arises in this setup even if the optical modes are tuned in resonance.

There are several ways to implement the phase shift between the classical and quantum components of the probing light. The most universal and straightforward, but not the most practical one is shown in Fig.\,\ref{fig:comb_simple}. Here the probe beam is
combined with an auxiliary one $d$ at the asymmetric beamsplitter with the amplitude transmissivity $T\to1$ and the reflectivity $R\to0$. In this case the quantum component just passes through the beamslitter unchanged, $\hat{a}_2^{c,s} = \hat{b}_1^{c,s}$. At the same time, suppose that the classical amplitude of the auxiliary beam $D$ is very strong: $RD\to\const$. In this case,
\begin{equation}
  A_2^{c,s} = B_1^{c,s} + RD^{c,s} \,.
\end{equation}
Evidently, any values of $A_2^{c,s}$ can be obtained in this way.

\begin{figure}
  \includegraphics{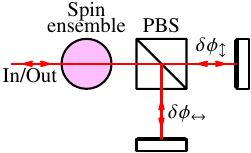}
  \caption{The double-pass scheme using an atomic spin ensemble. PBS: the polarization beamsplitter.}\label{fig:comb_spin}
\end{figure}

In the two important particular cases, namely the atomic spin systems and the Michelson interferometers (including the GW detectors), the specific features of the respective  systems could be used to create the phase shift. In the atomic spin system case, the Faraday interaction provides the coupling between the light and the atomic ensemble \cite{Moeller_Nature_547_191_2017}. This means that the classical and quantum components of the probing light belong to two orthogonal polarization. Therefore, after the first interaction they can be separated by the polarization beamsplitter and their phases can be adjusted independently as shown in Fig.\,\ref{fig:comb_spin}. Then they can be reflected back, combined together by the same polarization beamsplitter, and passed through the spin ensemble a second time.

It is worth noting that the scheme similar to the one shown in Fig.\,\ref{fig:comb_spin} can be used also for swapping of the quantum states between the light and spin ensemble, as well as for generation of entanglement between these susbsystems, see Ref.\,\cite{Muschik_PRA_73_062329_2006}.

\begin{figure}
  \includegraphics{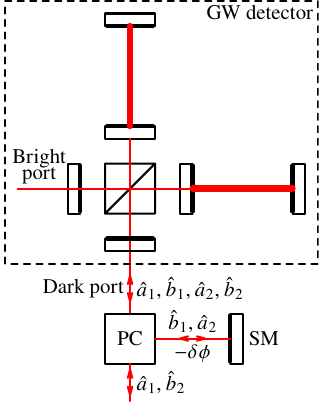}
  \caption{The double-pass GW detector. PC: the polarization circulator; SM: the steering mirror.}\label{fig:comb_GW}
\end{figure}

In the Michelson/Fabry-Perot topology of the GW detectors, see Fig.\,\ref{fig:comb_GW}, the classical component of the optical fields in the arms is coupled to the so-called ``bright'' input/output port of the interferometer, and the quantum component --- to another (``dark'') port. Therefore, the classical and quantum components can be manipulated independently, introducing the necessary phase shift between them.

Evidently, in this scheme the optical fields belonging to the first and the second pass through the interferometer has to be made distinguishable in some way (in particular, in the scheme of Fig.\,\ref{fig:double}, two cavities are used, and in the scheme of Fig.\,\ref{fig:comb_spin}, the first and the second passes correspond to the opposite propagation direction).

A possible method of ``coloring'' the first and second passes is the use of two orthogonal polarizations of the probing light. The simplified scheme of the polarization double interaction interferometer is shown in Fig.\,\ref{fig:comb_GW}. Here the additional block labeled ``PC'' (polarization circulator) modifies polarizations of the beams passing through it and reroute them as follows: the incident beam $a_1$ goes to the interferometer; the returning beam $b_1$ is redirected to the additional steering mirror; the reflected beam $a_2$ acquires the phase shift $-\delta\phi$ and then goes to the interferometer again; finally, the second returning beam $b_2$ exits the interferometer.

As it was mentioned already in the Introduction, this scheme is very similar to the two-pass quantum speedmeter interferometer proposed in Ref.\,\cite{18a1DaKnVoKhGrStHeHi}. This similarity opens some interesting possibilities, discussed briefly in the Conclusion.

\section{Robustness to optical losses}\label{sec:losses}

\subsection{Standard optomechanical setup}\label{sec:losses_1}

In order to compare the robustness of the virtual rigidity and the ``real'' optical springs to the optical losses, we consider two characteristic examples. The first one is the ``ordinary'' optomechanical setup using the free probe mass (as in the contemporary GW detectors). In this case, converting the free mass into the harmonic oscillator by means of the real optical spring or the virtual rigidity, it is possible to achieve a narrow band gain of sensitivity around the new mechanical resonance frequency. The second example that will be considered in the following subsection is the hybrid scheme using the auxiliary atomic spin ensemble \cite{18a1KhPo, 19a1ZePoKh}.

Here, we limit ourselves to the most important type of the optical losses, namely the output ones, including the photodetection inefficiency. We model them by means of an imaginary beamsplitters with the power transmissivities $\eta<1$, which is located at the output of the optical setup and which mixes the output field $\hat{d}$ of the latter one with the vacuum noise $\hat{g}$ \cite{Leonhardt_PRA_48_4598_1993}:
\begin{equation}\label{b_loss}
  \hat{d}^{c,s} = \sqrt{\eta}\,\hat{b}^{c,s} + \sqrt{1-\eta}\,\hat{g}^{c,s} \,.
\end{equation}
We assume that the sine quadrature $\hat{d}^s$ is measured by the homodyne detector.

Combining Eqs.\,\eqref{IO_11} and \eqref{b_loss}, we obtain that
\begin{equation}
  \hat{d}^s(\Omega)
  = \sqrt{\eta}\Upsilon_\varkappa\chi(\Omega)[f(\Omega) + \hat{f}_{\rm sum}(\Omega)] ,
\end{equation}
where
\begin{equation}\label{f^sum_varkappa}
  \hat{f}_{\rm sum} (\Omega)
  = \frac{\chi^{-1}(\Omega) + \varkappa}{\Upsilon_\varkappa}\hat{a}^s(\Omega)
    + \Upsilon_\varkappa\hat{a}^c(\Omega)
    + \frac{\epsilon\chi^{-1}(\Omega)}{Y_\varkappa}\hat{g}^s(\Omega)
\end{equation}
is the normalized sum quantum noise,
\begin{equation}
  \Upsilon_\varkappa = \Upsilon\cos\phi
\end{equation}
is the modified coupling factor,
\begin{equation}
  \varkappa = \frac{\Upsilon^2}{2}\sin2\phi = \Upsilon_\varkappa^2\tan\phi
\end{equation}
is the virtual rigidity, compare with Eqs.\,\eqref{Ups_eff}, \eqref{kappa}, and
\begin{gather}
  \epsilon = \sqrt{\frac{1-\eta}{\eta}} \,.
\end{gather}
Note that while the shot noise term in Eq.\,\eqref{f^sum_varkappa} (the first one) is modified by the factor $\varkappa$, this is not the case for the last one, originating from the optical losses. Therefore, these two terms can not be optimized simultaneously, degrading the sensitivity of this method.

Suppose for simplicity that all incident fields are in the ground state (as it was mentioned in Ref.\,\,\cite{12a1DaKh}, in the schemes that use the optical rigidity, sophisticated frequency-dependent squeezing should be used; its implementation could be problematic). In this case, the double-sided spectral densities of the quadratures $\hat{a}_{c,s}$ and $\hat{g}_{c,s}$ are equal to $1/2$. Therefore, the spectral density of $\hat{f}_{\rm sum}$ is equal to
\begin{equation}\label{S_varkappa}
  S(\Omega) = \frac{1}{2}\biggl\{
      \frac{|\chi^{-1}(\Omega) + \varkappa|^2}{\Upsilon_\varkappa^2}
      + \Upsilon_\varkappa^2 + \frac{\epsilon^2\chi^{-2}(\Omega)}{\Upsilon_\varkappa^2}
    \biggr\} .
\end{equation}
Optimizing it in $\Upsilon_\varkappa$, we obtain the following sensitivity limitation:
\begin{equation}
  S(\Omega) \ge \sqrt{|\chi^{-1}(\Omega) + \varkappa|^2 + \epsilon^2|\chi(\Omega)|^{-2}}
  \ge \epsilon S_{\rm SQL}(\Omega) \,.
\end{equation}

Let us consider now the optical rigidity case described by equations \eqref{IO_K}. Note that up to the replacements
\begin{subequations}\label{subst_kappa}
  \begin{gather}
    \Upsilon\to\Upsilon_\kappa \,, \\
    \chi^{-1}\to\chi^{-1}+\kappa \,,
  \end{gather}
\end{subequations}
they are identical to Eqs.\,\eqref{IO_11}. Therefore, we can reuse Eqs.\,\eqref{f^sum_varkappa}, \eqref{S_varkappa}, setting in these equations $\varkappa=0$ and taking into account the replacements \eqref{subst_kappa}. As a result, we obtain that
\begin{gather}
  \hat{f}_{\rm sum} (\Omega)
  = \frac{\chi^{-1}(\Omega) + \kappa}{\Upsilon_\kappa}
      [\hat{a}^s(\Omega) + \epsilon\hat{g}^s(\Omega)]
    + \Upsilon_\kappa\hat{a}^c(\Omega) \,, \label{f^sum_kappa} \\
  S(\Omega) = \frac{1}{2}\biggl\{
      \frac{|\chi^{-1}(\Omega) + \kappa|^2}{\Upsilon_\kappa^2}(1+\epsilon^2)
      + \Upsilon_\kappa^2
    \biggr\} .
\end{gather}
In this case, both the shot noise and the losses terms are modified in the unified way. Therefore, both of them can be canceled by $\kappa=-\chi^{-1}$. As a result, arbitrary high sensitivity can be achieved at the modified mechanical resonance frequency, albeit only in a narrow band.

\subsection{Hybrid scheme}\label{sec:hybrid}

\begin{figure}
  \includegraphics{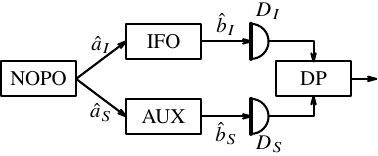}
  \caption{The hybrid scheme. NOPO: the non-degenerate optical parametric oscillator; IFO:  the interferometer, AUX: the auxiliary atomic spin system; $D_{I,A}$: the homodyne detector; DP: the data processing block. The subscripts ``$I$'' and ``$A$'' denote the interferometer and the auxiliary subsystems.}\label{fig:hybrid}
\end{figure}

The block diagram of the hybrid setup consisting of the optomechanical sensor IFO and the auxiliary system AUX is shown in Fig.\,\ref{fig:hybrid}. This topology was first discussed in Ref.\,\cite{Ma_NPhys_13_776_2017}; in that work, it was proposed to use some other optical mode of the optomechanical interferometer as the auxiliary system. Then in Refs.\,\cite{18a1KhPo, 19a1ZePoKh} implementations using the negative-frequency auxiliary spin oscillator were proposed.

Suppose that both subsystems are probed by the two entangled beams $\hat{a}_I$ and $\hat{a}_S$ generated by the non-degenerate optical parametric oscillator NOPO. The sine quadratures of the corresponding output beams $\hat{b}_I^s$, $\hat{b}_S^s$ are measured by the homodyne detectors and their output signals are combined by the data-processing block DP.  Here and below, the subscripts ``$I$'' and ``$S$'' stand for the ``Interferometer'' and ``Spin ensemble''.

We suppose that the optical rigidity, either virtual or real one, is used in the spin oscillator in order to align the susceptibilities $\chi_I$, $\chi_A$ of both subsystems. Also, in both cases, the effective optomechanical coupling factors are assumed to be matched. Therefore, in the virtual rigidity case,
\begin{subequations}
  \begin{gather}
    \chi_S^{-1}(\Omega) + \varkappa = -\chi_I(\Omega) \,, \label{con_virt_1} \\
    \Upsilon_{S \varkappa} = \Upsilon_I \,,
  \end{gather}
\end{subequations}
and in the real rigidity one,
\begin{subequations}
  \begin{gather}
    \chi_S^{-1}(\Omega) + \kappa = -\chi_I(\Omega) \,, \label{con_real_1} \\
    \Upsilon_{S \kappa} = \Upsilon_I \,.
  \end{gather}
\end{subequations}

The resulting sum noise spectral densities are calculated in App.\,\ref{app:S}. It is shown there that with account for some reasonable approximations, they can be presented as follows:
\begin{subequations}\label{S_sum_loss}
  \begin{equation}\label{S_sum_virt}
    S_{\rm sum}(\Omega) = \frac{1}{2\ch2r}
      \biggl[\frac{|\chi_I(\Omega)|^{-2}}{\Upsilon_I^2} + \Upsilon_I^2\biggr]
      + \frac{1}{2\Upsilon_I^2}
          [\epsilon_I^2|\chi_I(\Omega)|^{-2} + \epsilon_S^2|\chi_S(\Omega)|^{-2}]
  \end{equation}
  in the virtual rigidity case and
  \begin{equation}\label{S_sum_real}
    S_{\rm sum}(\Omega) = \frac{1}{2\ch2r}
      \biggl[\frac{|\chi_I(\Omega)|^{-2}}{\Upsilon_I^2} + \Upsilon_I^2\biggr]
      + \frac{\epsilon_I^2 + \epsilon_S^2}{2\Upsilon_I^2}|\chi_I(\Omega)|^{-2}
  \end{equation}
\end{subequations}
in the real rigidity one.

It follows from these equations that in the lossless case of $\epsilon_I=\epsilon_S=0$, both approaches give the same results, allowing to overcome the SQL \eqref{S_SQL} by the factor $\ch2r$. However, the terms proportional to the losses differ. They depend on the physical spin system susceptibility $\chi_s$ in the virtual rigidity case and on the modified one \eqref{con_virt_1} in the real rigidity one.

Consider the scenario discussed in the Introduction. Let the mechanical eigen frequency of the IFO subsystem is very low,
\begin{equation}
  \chi_I^{-1}(\Omega) \approx -\Omega^2 \,,
\end{equation}
while the one of the auxiliary channel it is higher than the characteristic mechanical force signal frequency:
\begin{gather}
  \chi_S^{-1}(\Omega) \approx \Omega^2 - \Omega_S^2 \,,\\
  \Omega \ll \Omega_S \,. \label{low_freq}
\end{gather}
Suppose that they are aligned using either virtual or real rigidity:
\begin{equation}
  \varkappa,\,\kappa = \Omega_S^2 \,.
\end{equation}
In this case, we obtain that in the low-frequency band of \eqref{low_freq},
\begin{subequations}\label{S_sum_loss_lf}
  \begin{equation}\label{S_sum_virt_2}
    S_{\rm sum}(\Omega)
    = \frac{1}{2\ch2r}\biggl(\frac{\Omega^4}{\Upsilon_I^2} + \Upsilon_I^2\biggr)
      + \frac{1}{2\Upsilon_I^2}(\epsilon_I^2\Omega^4 + \epsilon_S^2\Omega_S^4)\
  \end{equation}
  in the virtual rigidity case and
  \begin{equation}\label{S_sum_real_lf}
    S_{\rm sum}(\Omega)
    = \frac{1}{2\ch2r}\biggl(\frac{\Omega^4}{\Upsilon_I^2} + \Upsilon_I^2\biggr)
      + \frac{(\epsilon_I^2 + \epsilon_S^2)}{2\Upsilon_I^2}\,\Omega^4
  \end{equation}
\end{subequations}
in the real rigidity one. It is easy to see that in the latter case the term proportional to the spin ensemble channel inefficiency $\epsilon_S^2$ is $(\Omega_S/\Omega)^4$ times smaller.

\section{Conclusion}\label{sec:conclusion}

In this article, we propose a new type of the optical spring that does not require a detuned cavity and relies instead on the double interactions of the probing light with the mechanical object. Therefore, it is applicable in cases where high-finesse optical cavities can not be used --- in particular, in optomechanical setups using atomic spin oscillators \cite{18a1KhPo}.

Another advantage of this type of the optical spring is that the value of the corresponding spring constant can be adjusted easily by changing the phase shift introduced into the probing light between the interaction, without affecting the core optics of the optomechanical setup. This feature allows to vary the mechanical eigen frequency in a real time and therefore can be used in adaptive GW detection schemes, see \eg Ref.\,\cite{Simakov_PRD_90_102003_2014}.

The optical topology of the double-pass variant of the laser GW detectors topology  considered here (see Fig.\,\ref{fig:comb_GW}) is similar to the proposed in Ref.\,\cite{18a1DaKnVoKhGrStHeHi} double-pass implementation of the quantum speed meter concept \cite{90a1BrKh}, which allows to overcome the SQL by measuring the probe mass velocity instead of the position. The key difference is that in the quantum speedmeter case, the phase shift introduced between the two optomechanical interactions has to be equal to $2\psi=\pi$; in this case the optical spring vanishes, see Eq.\,\eqref{kappa}. This means that the scheme proposed here can be switched in a real time between the speed meter mode and the position meter one with the tunable optical spring. This possibility could be useful for the detection of GW signals from binary systems mergers.

\acknowledgments

This work was supported by the Theoretical Physics and Mathematics Advancement Foundation ``BASIS'' Grant \#23-1-1-39-1. The paper has been assigned LIGO document number P2400002.

The author would like to thank E.\,Polzik, E.\,Zeuthen, A.\,Markowitz and S.\,Danilishin for commenting on the manuscript.

\appendix

\section{Calculation of of the sum noise spectral densities of the hybrid scheme}\label{app:S}

For brevity, we suppress in the Appendix the explicit dependence on $\Omega$.

Reusing Eq.\,\eqref{f^sum_varkappa}, we obtain the following equations for the sum force-normalized noise of the interferometer and spin oscillator:
\begin{subequations}
  \begin{gather}
    \hat{f}_{I\,\rm sum}
      = \frac{\chi_I^{-1}}{\Upsilon_I}\hat{a}_I^s
        + \Upsilon_I\hat{a}_I^c
        + \frac{\epsilon_I\chi_I^{-1}}{\Upsilon_I}\hat{g}_I^s \,, \label{f_I} \\
    \hat{f}_{S\,\rm sum}
      = \frac{\chi_S^{-1} + k}{\Upsilon_I}\hat{a}_S^s + \Upsilon_I\hat{a}_S^c
        + \frac{\epsilon_S(\chi_S^{-1}+k_{\rm loss})}{\Upsilon_I}\hat{g}_S^s \,. \label{f_S}
  \end{gather}
\end{subequations}
where
\begin{subequations}
  \begin{gather}
    k = k_{\rm loss} = \kappa\quad\text{in the real rigidity case,}\\
    k = \varkappa\,,\ k_{\rm loss} = 0 \quad\text{in the virtual rigidity one.}
  \end{gather}
\end{subequations}
Spectral densities of the quadratures $\hat{a}_{I,S}^{c,s}$, and their non-vanishing cross spectral densities are equal to
\begin{subequations}
  \begin{gather}
    S^c_I = S^s_I = S^c_S = S^s_S = \frac{\ch2r}{2} \,, \\
    S^c_{IS} = \frac{\sh2r}{2} \,,\quad S^s_{IS} = -\frac{\sh2r}{2} \,.
  \end{gather}
\end{subequations}
Therefore, spectral densities of $\hat{f}_{I\,\rm sum}$, $\hat{f}_{S\,\rm sum}$, and their cross spectral density are equal to:
\begin{subequations}
  \begin{gather}
    S_I = S_{I0}\ch2r + S_{I\,{\rm loss}} \,, \\
    S_S = S_{S0}\ch2r + S_{S\,{\rm loss}} \,, \\
    S_{IS} = S_{IS0}\sh2r \,,
  \end{gather}
\end{subequations}
where
\begin{subequations}
  \begin{gather}
    S_{I0} = \frac{1}{2}
      \biggl[\frac{|\chi_I|^{-2}}{\Upsilon_I^2} + \Upsilon_I^2\biggr] , \\
    S_{S0} = \frac{1}{2}
      \biggl[\frac{|\chi_S^{-1}+k|^2}{\Upsilon_I^2} + \Upsilon_I^2\biggr],\\
    S_{SI0} = \frac{1}{2}\biggl[
        \frac{|\chi_I|^{-1}|\chi_S^{-1}+k|}{\Upsilon_I^2} - \Upsilon_I^2
      \biggr] , \\
    S_{I\,{\rm loss}} = \frac{\epsilon_I^2|\chi_I|^{-2}}{2\Upsilon_I^2} \,, \\
    S_{S\,{\rm loss}} = \frac{\epsilon_S^2|chi_S^{-1}+k_{\rm loss}|^2}{2\Upsilon_I^2} \,.
  \end{gather}
\end{subequations}

Taking into account that the ``$I$'' channel contains the signal, the optimized noise spectral density at the output of the data processing unit is equal to
\begin{multline}\label{S^sum_2}
  S_{\rm sum} = S_IS - \frac{S_{IS}^2}{S_S} \\
  = \frac{1}{S_S}\bigl\{
         [S_{S0}S_{I0} - S_{SI0}^2]\sh^2r +  S_{S0}S_{I0}
         + [S_{S0}S_{I\,{\rm loss}} + S_{I0}S_{S\,{\rm loss}}]\ch2r
         + S_{I\,{\rm loss}}S_{S\,{\rm loss}}
      \bigr\} \,.
\end{multline}
It follows from Eq.\,\eqref{con_virt_1} that
\begin{subequations}
  \begin{gather}
    S_{I0}S_{S0} - S_{SI0}^2 = 0 \,, \\
    S_{S0} = S_{I0} \,,
  \end{gather}
\end{subequations}
and
\begin{equation}
  S_{\rm sum} = \frac{
      S_{I0}^2 + S_{I0}(S_{I\,{\rm loss}} + S_{S\,{\rm loss}})\ch2r
      + S_{I\,{\rm loss}}S_{S\,{\rm loss}}
    }{S_{I0}\ch2r + S_{S\,{\rm loss}}} \,.
\end{equation}
Neglecting the small terms $S_{I\,{\rm loss}}S_{S\,{\rm loss}}$ in the numerator and $S_{S\,{\rm loss}}$ in the denominator, we obtain Eqs.\,\eqref{S_sum_loss}.


%

\end{document}